\documentclass[%
 amsmath,amssymb,
 reprint,%
]{revtex4-1}

\usepackage{subfigure}
\usepackage{graphicx}% Include figure files
\usepackage{dcolumn}% Align table columns on decimal point
\usepackage{bm}% bold math
\usepackage{color}
\begin{document}

\preprint{AIP/123-QED}

\title{Effects of reciprocity on random walks in weighted networks}%: A spectrum based approach
% Force line breaks with \\
%\thanks{Footnote to title of article.}

\author{Zhongzhi Zhang}
\email{zhangzz@fudan.edu.cn}
%\homepage{http://www.researcherid.com/rid/G-5522-2011}

\author{Huan Li}

\author{Yibin Sheng}

\affiliation {School of Computer Science, Fudan University,
Shanghai 200433, China}

\affiliation {Shanghai Key Lab of Intelligent Information
Processing, Fudan University, Shanghai 200433, China}

\date{\today}% It is always \today, today,

\begin{abstract}
It has been recently reported that the reciprocity of real-life weighted networks is very pronounced, however its impact on dynamical processes is poorly understood. In this paper, we study random walks in a scale-free directed weighted network with a trap at the central hub node, where the weight of each directed edge is dominated by a parameter controlling the extent of network reciprocity. We derive the mean first passage time (MFPT) to the trap, by using two different techniques, the results of which agree well with each other. We also analytically determine all the eigenvalues as well as their multiplicities for the fundamental matrix of the dynamical process, and show that the largest eigenvalue has an identical dominant scaling as that of the MFPT. We find that the weight parameter has a substantial effect on the MFPT, which behaves as a power-law function of the system size with the power exponent dependent on the parameter, signaling the crucial role of reciprocity in random walks occurring in weighted networks.
\end{abstract}

%05.40.Fb Random walks and Levy flights
%61.43.Hv  Fractals; macroscopic aggregates (including diffusion-limited aggregates)
%05.45.Df Fractals
%05.60.Cd Classical transport
%05.40.-a Fluctuation phenomena, random processes, noise, and Brownian motion
%89.75.Hc Networks and genealogical trees
%36.20.�Cr Macromolecules and polymer molecules

\maketitle

%\tableofcontents

\section{Introduction}

As an emerging science, complex networks have witnessed substantial progress in the past years~\cite{Ne03}. One of the ultimate goals in the study of complex networks is to uncover the influences of various structural properties on the function or dynamical processes taking place on them. Among different dynamical processes, random walks lie at the core, since they are a fundamental mechanism for a wealth of other dynamic processes, such as navigation~\cite{Kl00}, search~\cite{GuDiVeCaAr02,BeLoMoVo11}, and cooperative control~\cite{Ol07}. Except for the importance in the area of network science, random walks also provide a paradigmatic model for analyzing and understanding a large variety of real-world phenomena, for example, animal~\cite{BadaViCa05} and human~\cite{BrHuGe06} mobility. Thus far, random walks have found numerous applications~\cite{We05} in many aspects of interdisciplinary sciences, including image segmentation~\cite{Le06}, community detection~\cite{PoLa06,RoEsLaWeLa14}, collaborative recommendation~\cite{FoPiReSa07}, and signal propagation in proteins~\cite{ChBa07} to name a few.

A highly desirable quantity for random walks is first passage time (FPT)~\cite{Re01}, defined as the expected time for a random walker going from a starting node to a given target. The mean of FPTs over all starting nodes to the target is called mean first passage time (MFPT), which is an important characteristic of random walks due to the first encounter properties in numerous realistic situations. In the past years, the study of MFPT has triggered an increasing attention from the scientific community~\cite{BeVo14,LiZh14}. One focus of theoretical activity is to develop general methods to efficiently compute MFPT~\cite{NoRi04,BeCoMo05,CoBeKl07,CoBeTeVoKl07,CoBeTeVoKl07}. Another direction is to unveil how the behavior of MFPT is affected by different structural properties of the underlying systems, such as heterogeneity of degree~\cite{ZhQiZhXiGu09} or strength~\cite{LiZh13}, fractality~\cite{ZhXiZhGaGu09}, and modularity~\cite{ZhLiGaZhGu09}.

Previous studies proposed several frameworks for evaluating MFPT and uncovered the discernible effects of some nontrivial structural aspects on the target search efficiency measured by MFPT. However, most existing works ignoring the impact of link reciprocity, the tendency of node pairs to form mutual connection in directed networks, on the behavior of random walks, despite the fact that reciprocity is a common characteristic of many realistic networks~\cite{GaLo04}, such as the World Wide Web~\cite{AlJeBa99}, e-mail networks~\cite{EbMiBo02,NeFoBa02}, and World Trade Web~\cite{SeBo03}. In addition to binary networks, the nontrivial pattern reciprocity is also ubiquitous in real-life systems described by weighted networks~\cite{AkVaFa12,WaLiHaStToCh13,SqPiRuGa13}. It has been shown the ubiquitous link reciprocity strongly affects dynamical processes in binary networks, for example, spread of computer viruses~\cite{NeFoBa02} or information~\cite{ZhZhSuTaZhZh14}, and percolation~\cite{BoSa05}. By contrast, the influence of reciprocity on dynamical processes in weighted networks has attracted much less attention, although it is suggested that reciprocity could play a crucial role in network dynamics. In particular, the lack of analytical results in this field limits our understanding of the impact of weight reciprocity on the function of weighted networks~\cite{SqPiRuGa13}.

In this paper, we propose a weighted directed scale-free network by replacing each edge in the previous binary network~\cite{SoHaMa06,RoHaBe07} by double links with opposite directions and different weights. In the weighted network, the link weights are adjusted by a parameter characterizing the weight reciprocity of network. We then study random walks in the weighted network in the presence of a perfect trap at the central large-degree node. During the process of random walks, the transition probability is dependent on the weight parameter. We derive the MFPT to the target by using two disparate approaches, the results of which completely agree with each other. We also determine all the eigenvalues and their multiplicities of the fundamental matrix characterizing the random-walk process, and show that the largest eigenvalue has the same leading scaling as that of the MFPT. The obtained results demonstrate that the behavior of MFPT to the trap depends on the weighted parameter, signalling a drastic influence of the  weight reciprocity on random walks defining on weighted networks.

%%%%%%%%%%%%%%%%%%%%%%%%%%%%%%%%%%%%%%%%%%%%%%%%%%%%%%%%%
% Figure  1
%%%%%%%%%%%%%%%%%%%%%%%%%%%%%%%%%%%%%%%%%%%%%%%%%%%%%%%%%%
\begin{figure}
\begin{center}
\includegraphics[width=0.80\linewidth]{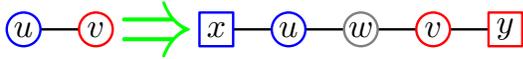}
\caption{Illustration of construction of the binary network. The next generation is obtained from current generation by replacing each edge with the cluster on the right-hand side of the arrow, where $w$ is
new internal node, while $x$ and $y$ are external nodes.}
\label{cons}
\end{center}
\end{figure}
%%%%%%%%%%%%%%%%%%%%%%%%%%%%%%%%%%%%%%%%%%%%%%%%%%%%%%%%%%

\section{Network models and properties\label{model}}

Before introducing the weighted directed network with scale-free fractal properties. We first give a brief introduction to a binary scale-free fractal network, which has the same topology as the weighted network.

\subsection{Model and properties of binary network.}

\begin{figure}
\begin{center}
\includegraphics[width=0.95\linewidth]{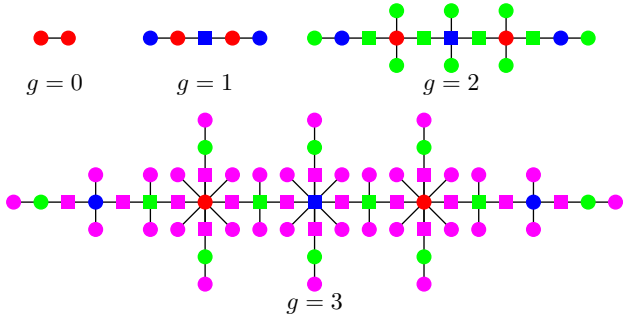}
\end{center}
\caption{Iterative growth processes for the first several generations.}
\label{network}
\end{figure}

The binary treelike network is constructed in an iterative way~\cite{SoHaMa06,RoHaBe07}. Let $F_{g}$ ($g \ge 0$) represent the network after $g$ iterations (generations). For $g=0$, $F_{0}$ is an edge linked by two nodes. In each successive iteration $g \ge 1$, $F_{g}$ is
constructed from $F_{g-1}$ by performing the following operations on every existing edge in $F_{g-1}$ as shown in Fig.~\ref{cons}: two new nodes (called external
nodes) are firstly created and attached,
respectively, to both  endpoints of the edge; then, the edge is
broken, another new node (referred to as an internal node) is placed in its middle and linked to both endpoints of the original edge. Figure~\ref{network} illustrates the first several construction processes of the network. The structure of $F_{g}$ is enciphered in its adjacency matrix $A_g$, the entries $A_g(i,j)$ of which are defined by $A_g(i,j)=1$ if two nodes $i$ and $j$ are adjacent in $F_{g}$, or $A_g(i,j)=0$ otherwise.

%%%%%%%%%%%%%%%%%%%%%%%%%%%%%%%%%%%%%%%%%%%%%%%%%%%%%%%%%
% Figure  2
%%%%%%%%%%%%%%%%%%%%%%%%%%%%%%%%%%%%%%%%%%%%%%%%%%%%%%%%%%
%%%%%%%%%%%%%%%%%%%%%%%%%%%%%%%%%%%%%%%%%%%%%%%%%%%%%%%%%%

The particular construction of the network allows to calculate exactly its relevant properties. At each generation $g_i$ ($g_i\geq 1$), the number of newly created nodes is $\Upsilon_{g_i}=3\cdot4^{g_i-1}$. Let $\overline{\Lambda}_{g_i}$ be the set of nodes generated at iteration $g_i$, then $\overline{\Lambda}_{g_i}$ can be further classified
into two sets $\overline{\Lambda}_{g_i, {\rm ext}}$ and $\overline{\Lambda}_{g_i, {\rm
int}}$ satisfying $\overline{\Lambda}_{g_i} =
\overline{\Lambda}_{g_i, {\rm ext}} \cup \overline{\Lambda}_{g_i, {\rm
int}}$, among which  $\overline{\Lambda}_{g_i, {\rm ext}}$  is the set of external nodes and
$\overline{\Lambda}_{g_i, {\rm
int}}$ is the set of internal nodes. We use $|\Omega|$ to stand for the cardinality of a set $\Omega$. Because $|\overline{\Lambda}_{g_i, {\rm ext}}|
=2|\overline{\Lambda}_{g_i, {\rm in}}|$, it is easy to derive  $|\overline{\Lambda}_{g_i, {\rm int}}| = 4^{g_i-1}$ and
$|\overline{\Lambda}_{g_i, {\rm ext}}| = 2\cdot4^{g_i-1}$. We represent the set of nodes in
$F_g$ as $\Lambda_g$. Hence, the number of nodes and edges in $F_g$ is $N_g=|\Lambda_g|=\sum_{g_i=0}^{g}\Upsilon_{g_i}=4^{g}+1$ and $E_g=N_g-1=4^{g}$, respectively.
Let $k_i(g)$ denote the degree of an arbitrary node $i$ in $F_g$ that was generated at generation $g_i$ ($g_i\geq 0$), then $k_i(g+1)=2\,k_i(g)$. Hence, after each new iteration the degree of every node doubles.

This resultant network displays the remarkable scale-free~\cite{BaAl99} and fractal~\cite{SoHaMa05} features as observed in diverse real-life systems. It has a power law degree distribution with an exponent $3$, and its fractal dimension is $2$.

\subsection{Model and properties of weighted directed network.}

The above introduced binary network $F_{g}$ can be extended to a weighted directed network with nonnegative and asymmetrical edge weights.
Let $\vec{F}_{g}$ denote the weighted directed network corresponding to $F_{g}$. Both  $\vec{F}_{g}$ and $F_{g}$ have an identical topological structure. The only difference between $\vec{F}_{g}$ and $F_{g}$ is that every undirected  edge in $F_{g}$ is replaced by two directed edges with opposite directions and distinct positive weights. We use
$W_g$ to represent the nonnegative and asymmetrical weight matrix for $\vec{F}_{g}$ such that
$W_{ij}(g)>0$ if and only if there is a directed edge (arc) pointing to node $j$ from node $i$. The weight
of each arc in the weighted directed network is defined recursively in the following way. When $g=0$,
$\vec{F}_{0}$ has two nodes, denoted by $a$ and $b$, and the weights of arcs $\vec{e}(a,b)$ and
$\vec{e}(b,a)$ are defined to be $W_{ab}(0)=W_{ba}(0)=1$. When $g \geq 1$, by construction, $F_g$
is obtained from $F_{g-1}$ by substituting each undirected edge $e(u,v)$ in $F_{g-1}$ with
two undirected edges $e(u,w)$ and $e(w,v)$, and generating two additional nodes, $x$ and
$y$, attaching to $u$ and
$v$, respectively. The weights of resultant arcs in $\vec{F}_{g}$ are
defined as: $W_{uw}(g)=W_{uv}(g-1)$, $W_{vw}(g)=W_{vu}(g-1)$, $W_{wu}(g)=W_{wv}(g)=1$,
$W_{xu}(g)=1$, $W_{yv}(g)=1$,
$W_{ux}(g)=\theta \, W_{uv}(g-1)$, and
$W_{vy}(g)=\theta\, W_{vu}(g-1)$.
Here $\theta$ is a tunable positive real number, that is, $\theta>0$. The weight parameter is of paramount importance since it characterizes the weight reciprocity of network $\vec{F}_{g}$. When  $\theta=1$, $\vec{F}_{g}$ reduces to $F_{g}$, and the weights in two directions between any pair of adjacent nodes are completely reciprocated; when $\theta \neq 1$, the weights are non-reciprocated: the larger the deviation of $\theta$ from 1, the smaller the level of weight reciprocity.

In undirected weighted networks~\cite{BaBaPaVe04}, node strength is a key quantity
characterizing the property of a node. Here we extend the definition of strength of a node  to the directed weighted network $\vec{F}_{g}$ by defining the out-strength and in-strength
of node $i$ in $\vec{F}_{g}$ as $s_i^+(g)=\sum_{j=1}^{N_g}W_{ij}(g)$ and
$s_i^-(g)=\sum_{j=1}^{N_g}W_{ji}(g)$, respectively.
For $\vec{F}_{g}$, we can obtain the out-strength for an arbitrary node $i$
that entered the network at generation $g_i$ ($g_i\geq 0$). If $i$ was an external node when it entered the network, $s_i^+(g)=(\theta+1)^{g-g_{i}}$;
otherwise, if $i$ was an internal node when it was born, $s_i^+(g)=2(\theta+1)^{g-g_{i}}$.
Therefore, after each new iteration, the out-strength of a node increases by a factor of $\theta$. It is easy to obtain the node out-strength in $\vec{F}_{g}$ obeys a distribution of  power law form with the exponent being $1+\frac{2\ln 2}{\ln (\theta+1)}$. Note that in some realistic networks, the node strength also display a broad distribution~\cite{BaBaPaVe04}.

\section{Formulation of biased walks in the weighted directed network}

After introducing the construction and property of the weighted directed network $\vec{F}_{g}$, we now define and study biased discrete-time random walks performing $\vec{F}_{g}$. Let $r_{ij}(g)=W_{ij}(g)/s_i^+(g)$ denote the transition probability that a particle jumps from node $i$ to its neighboring node $j$ per time step. Note that $r_{ij}(g)$ constitutes an entry of transition matrix $R_g=(S_g)^{-1}W_g$, where $S_g$ is the diagonal
out-strength matrix of $\vec{F}_{g}$, with the $i$th diagonal entry of $S_g$ being $s_i^+(g)$.

In this paper, we focus on a specific case of biased random walks, often called trapping problem, in $\vec{F}_{g}$ in the presence of a trap placed at the central hub node, i.e., the internal node generated at the first iteration. To facilitate the description of the following text, all $N_g$ nodes in $\vec{F}_{g}$ are labeled sequentially as $1,2,\ldots, N_{g} -1,  N_g$ as follows. For
$\vec{F}_{1}$ , the newly generated internal node is labeled 1, the
initial two nodes in $\vec{F}_{0}$ are labeled as 2 and
3, while the two new external nodes are labeled by 4 and 5. For each new iteration $g_i >1$, we label consecutively the new nodes born at this iteration from $N_{g_i-1} + 1$ to $N_{g_i}$, \ while we keep the labels of those nodes created before iteration $g_i$ unchanged.

For the trapping problem, what we are concerned with are the trapping time and the average trapping time. Let $T_i^{(g)}$ represent the trapping time for a particle initially placed at node $i$ ($i \neq 1$) in $\vec{F}_{g}$ to arrive at the trap node for the first time, which is equal to the FPT from the $i$ to the trap. The average trapping time, $\langle T \rangle_g$, is actually the MFPT to the trap, defined as the mean of $T_i^{(g)}$ over all non-trap initial nodes in network $F_g$:
\begin{equation}\label{FPT}
\langle T \rangle_g=\frac{1}{N_g-1}\sum_{i=2}^{N_g} T_i^{(g)}\,.
\end{equation}
Below we will show how to compute  the two quantities $T_i^{(g)}$ and $\langle T \rangle_g$.

For $T_i^{(g)}$, it obeys the  relation
\begin{equation}\label{Form01}
T_i^{(g)}=1 +\sum_{j=2}^{N_g} r_{ij} T_j^{(g)}\,,
\end{equation}
which can be recast in matrix form as:
\begin{equation}\label{Form02}
T = e + \bar{R}_g \,T \,,
\end{equation}
where $T = \left(T_2^{(g)},T_3^{(g)},\ldots,T_{N_g}^{(g)}\right)^{\top}$ is an $(N_g-1)$-dimensional vector, $e=(1,1,...,1)^{\top}$ is the $(N_g-1)$-dimensional vector of all ones, and $\bar{R}_g$ is a matrix of order $N_g-1$, which a submatrix of $R_g$ and obtained from  $R_g$ by deleting the first row and the first column corresponding to the trap. From Eq.~(\ref{Form02}) we have
\begin{equation}\label{Form03}
T=(I-\bar{R}_g)^{-1}e=K_g\,e\,,
\end{equation}
where $I$ is the $(N_g-1)\times(N_g-1)$ identity matrix. Matrix $K_g=(I-\bar{R}_g)^{-1}$ is the fundamental matrix~\cite{KeSn60} of the addressed trapping problem. Equation~(\ref{Form03}) implies
\begin{equation}\label{Form04}
T_i^{(g)}=\sum_{j=2}^{N_g} K_g(i,j)\,,
\end{equation}
where $K_g(i,j)$ is the $ij$th entry of  matrix $K_g$, representing the expected number of visitations to node $j$ by a particle starting from node $i$ before being
absorbed by the trap.
Plugging Eq.~(\ref{Form04}) into Eq.~(\ref{FPT}) yields
\begin{equation}\label{Form05}
\langle T \rangle_g=\frac{1}{N_g-1}\sum_{i=2}^{N_g}\sum_{j=2}^{N_g} K_g(i,j)\,.
\end{equation}

Equation~(\ref{Form05}) indicates that the computation of MFPT $\langle T \rangle_g$ can be reduced to finding the sum of all entries of the corresponding fundamental matrix. A disadvantage  of this method is that it demands a large computational effort when the network is very large.
However, Eq.~(\ref{Form05}) provides exact results for $\langle T \rangle_g$ that can be applied to check the results for MFPT obtained using other techniques.
Next we analytically determine the closed-form expression for MFPT $\langle T \rangle_g$ using an alternative approach, the results of which are consistent with those of Eq.~(\ref{Form05}).

\section{Exact solution to the MFPT $\langle T \rangle_g$}

The particular selection of trap location and the specific network structure allow to determine exactly the MFPT $\langle T \rangle_g$ for arbitrary $g$.
In order to obtain a close-form expression for $\langle T \rangle_g$, we first establish the dependence of $T_i^{(g)}$ on iteration $g$.
For a node $i$ in  $\vec{F}_{g}$, at iteration $g+1$, its degree doubles, increasing
from $k_i(g)$ to $2k_i(g)$. All these $2k_i(g)$
neighboring nodes are created at iteration $g+1$, among which
one half are external nodes with a single degree, and the other half are internal nodes with degree 2.

We now consider the trapping problem in $\vec{F}_{g+1}$. Let $A$ be the FPT for a particle starting from node $i$ to any
of its $k_i(g)$ old neighbors, that is, those nodes adjacent to $i$ at iteration $g$; let $B$ (resp. $C$) be the
FPT for a particle staring from any of the $k_i(g)$ internal (resp. external) neighbors of $i$ to one of its $k_i(g)$ old neighbors. Then the FPTs obey relations:
\begin{eqnarray}\label{FPTF1}
\left\{
\begin{array}{ccc}
A&=&\frac{\theta}{\theta+1}(1+C) + \frac{1}{\theta+1}(1+B)\,,\\
B&=&\frac{1}{2} + \frac{1}{2}(1+A)\,,\\
C&=&1+A\,.
 \end{array}
 \right.
\end{eqnarray}
Eliminating $B$ and $C$ in Eq.~(\ref{FPTF1}), we obtain $A=4(\theta+1)$.
Therefore, when the network grows from iteration $g$ to iteration
$g+1$, the FPT from any node $i$ ($i \in \vec{F}_g$) to another node
$j$ ($j \in \vec{F}_g$) increases by a factor of $4(\theta+1)$. Hence,  $T_i^{(g+1)}=4(\theta+1)\,T_i^{(g)}$ hold for any $g$, which is a useful for
deriving the exact expression for MFPT.

Having obtained the scaling dominating the evolution for FPTs, we continue 
determining the MFPT $\langle T \rangle_g$. For this purpose,
we introduce two intermediary quantities for any $n \leq g$:
$T_{n,{\rm tot}}^{(g)}=\sum_{i\in \Lambda_n} T_i^{(g)}$ and
$\overline{T}_{n,{\rm tot}}^{(g)}=\sum_{i\in \overline{\Lambda}_n}
T_i^{(g)}$. Then,
\begin{equation}\label{FPTF2}
T_{g, {\rm tot}}^{(g)} = T_{g - 1, {\rm tot}}^{(g)} +\overline{T}_{g, {\rm tot}}^{(g)}= (4\theta+4)\,T_{g - 1, {\rm tot}}^{(g-1)} + \overline{T}_{g, {\rm tot}}^{(g)}\,.
\end{equation}
By definition, $\langle T \rangle_g=\frac{1}{N_g-1}T_{g, {\rm tot}}^{(g)}$. To find $T_{g, {\rm tot}}^{(g)}$, it is necessary to explicitly determine the quantity $\overline{T}_{g, {\rm
tot}}^{(g)}$. To this end, we define two additional
quantities for $n \leq g$: $\overline{T}_{n, {\rm int}}^{(g)}
= \sum_{i \in \overline{\Lambda}_{n, {\rm int}}}T_i^{(g)}$ and
$\overline{T}_{n, {\rm ext}}^{(g)} = \sum_{i \in
\overline{\Lambda}_{n, {\rm ext}}} T_i^{(g)}$. Obviously,
$\overline{T}_{g, {\rm tot}}^{(g)} = \overline{T}_{g, {\rm int}}^{(g)} + \overline{T}_{g, {\rm
    ext}}^{(g)}$. Thus, in order to find $\overline{T}_{g, {\rm tot}}^{(g)}$, one
may alternatively evaluate $\overline{T}_{n, {\rm int}}^{(g)}$ and
$\overline{T}_{n, {\rm ext}}^{(g)}$.

We first establish the relationship between $\overline{T}_{n, {\rm int}}^{(g)}$ and $\overline{T}_{n, {\rm
ext}}^{(g)}$. By construction (see Fig.~\ref{cons}), at a given
generation, each edge connecting two nodes $u$ and $v$  will give rise three new nodes ($w$, $x$, and $y$) in the next
generation. The two external nodes $x$ and $y$
are separately attached to $u$ and $v$, while the only internal node
$w$ is linked simultaneously to $u$ and $v$.
For any iteration $g$, the FPTs for the three new nodes satisfy:
$T^{(g)}_x=1 +T^{(g)}_u$, $T^{(g)}_y=1 +T^{(g)}_v$, and $T^{(g)}_w=\frac{1}{2}\left[1+T^{(g)}_u\right]+\frac{1}{2}\left[1+T^{(g)}_v\right]$. Therefore,
$T^{(g)}_x+T^{(g)}_y = 2\,T^{(g)}_w$.
Summing this relation over all old edges at the generation
before growth, we find that for all $n \leq g$, $\overline{T}_{n, {\rm ext}}^{(g)} = 2\,\overline{T}_{n, {\rm int}}^{(g)}$ always holds. In this way,
issue of determining $\overline{T}_{g, {\rm tot}}^{(g)}$ is reduced to finding $\overline{T}_{g, {\rm ext}}^{(g)}$ that can
be obtained as follows.

For an arbitrary external node $i_{\rm ext}$ in $\vec{F}_g$,
which is created at generation $g$ and attached to an old node $i$,
we have $T_{i_{\rm ext}}^{(g)} = 1 + T_i^{(g)}$, a relation valid for any node
pair containing an old node and one of its new external adjacent
nodes. By applying relation $T_{i_{\rm ext}}^{(g)} = 1 + T_i^{(g)}$ to two sum (the first one is
over a given old node and all its new adjacent external nodes, the other is
summing the first one over all old nodes), we obtain
\begin{eqnarray}\label{FPTF9}
\overline{T}_{g, {\rm ext}}^{(g)} &=&|\overline{\Lambda}_{g, {\rm
ext}}|+\sum_{i \in \Lambda_{g-1}}\left(k_i(g-1)\times
T_i^{(g)}\right) \nonumber\\
&=&|\overline{\Lambda}_{g, {\rm ext}}|+\left(\overline{T}_{g-1, {\rm ext}}^{(g)}+2\overline{T}_{g-1, {\rm int}}^{(g)}\right)
\nonumber\\
&\quad& +\left(2\overline{T}_{g-2, {\rm ext}}^{(g)}+4\overline{T}_{g-2, {\rm int}}^{(g)}\right) + \cdots \nonumber\\
&\quad& +\left(2^{g-2}\overline{T}_{1, {\rm ext}}^{(g)}+2^{g-1}\overline{T}_{1, {\rm int}}^{(g)}\right) \nonumber\\
&=& 2\times4^{g-1}+2\overline{T}_{g-1, {\rm ext}}^{(g)}+4\overline{T}_{g-2, {\rm ext}}^{(g)}+\cdots \nonumber\\
&\quad& +2^{g-1}\overline{T}_{1, {\rm ext}}^{(g)}
\end{eqnarray}
%where Eqs.~(\ref{FPTF4}) and~(\ref{FPTF7}) were used.

From Eq.~(\ref{FPTF9}), one can derive the recursive relation
\begin{eqnarray} \label{FPTF13}
    \overline{T}_{g+1, {\rm ext}}^{(g+1)} &=& 16(\theta+1)\,\overline{T}_{g, {\rm ext}}^{(g)} - (4\theta+2)4^g\,.
\end{eqnarray}
Considering the initial condition $\overline{T}_{2, {\rm ext}}^{(2)}
= 48\theta^2+80\theta+40$, Eq.~(\ref{FPTF13}) is solved
to yield
\begin{eqnarray} \label{FPTF14}
\overline{T}_{g, {\rm ext}}^{(g)} &=& \frac{12\theta^2+17\theta+7}{(\theta+1)(4\theta+3)}2^{4g-4}(\theta+1)^g + \frac{2\theta+1}{4\theta+3}2^{2g-1}\,.\notag\\
\end{eqnarray}
Because $\overline{T}_{g, {\rm tot}}^{(g)} = \overline{T}_{g, {\rm int}}^{(g)} + \overline{T}_{g, {\rm ext}}^{(g)}$ and $\overline{T}_{g, {\rm ext}}^{(g)}=2\,
\overline{T}_{g, {\rm int}}^{(g)}$, we have
\begin{equation} \label{FPTF15}
\overline{T}_{g, {\rm
tot}}^{(g)}= \frac{36\theta^2+51\theta+21}{(\theta+1)(4\theta+3)}2^{4g-5}(\theta+1)^g +\frac{6\theta+3}{4\theta+3}2^{2g-2}\,.
\end{equation}
Inserting Eq.~(\ref{FPTF15}) into Eq.~(\ref{FPTF2}) leads to
\begin{eqnarray}\label{FPTF16}
T_{g, {\rm tot}}^{(g)} &=& (4\theta+4)\,T_{g-1, {\rm tot}}^{(g-1)}+\frac{6\theta+3}{4\theta+3}2^{2g-2} \notag\\
 &\quad& +\frac{36\theta^2+51\theta+21}{(\theta+1)(4\theta+3)}2^{4g-5}(\theta+1)^g\,.
\end{eqnarray}
Using $T_{1, {\rm tot}}^{(1)} = 8\theta+6$, Eq.~(\ref{FPTF16})
is solved to get
\begin{eqnarray}\label{FPTF17}
T_{g, {\rm tot}}^{(g)} &=& \frac{12\theta^3+17\theta^2+7\theta}{\theta(\theta+1)(4\theta+3)}2^{4g-3}(\theta+1)^g  \notag\\
&\quad& +\frac{16\theta^3+28\theta^2+20\theta+6}{\theta(\theta+1)(4\theta+3)}2^{2g-3}(\theta+1)^g \notag\\
&\quad& -\frac{3(\theta+1)(2\theta+1)}{\theta(\theta+1)(4\theta+3)}2^{2g-2}\,.
\end{eqnarray}
Then, the rigorous expression for the MFPT $ \langle T
\rangle_g$ of the weighted directed network $\vec{F}_g$ is
\begin{eqnarray}\label{FPTF18}
 \langle T
\rangle_g
&=& \frac{12\theta^3+17\theta^2+7\theta}{8\theta(\theta+1)(4\theta+3)}2^{2g}(\theta+1)^g  \notag\\
&\quad& +\frac{16\theta^3+28\theta^2+20\theta+6}{8\theta(\theta+1)(4\theta+3)}(\theta+1)^g \notag\\
&\quad& -\frac{3(\theta+1)(2\theta+1)}{4\theta(\theta+1)(4\theta+3)}\,.
\end{eqnarray}

\begin{figure}
\begin{center}
\includegraphics[width=1.0\linewidth]{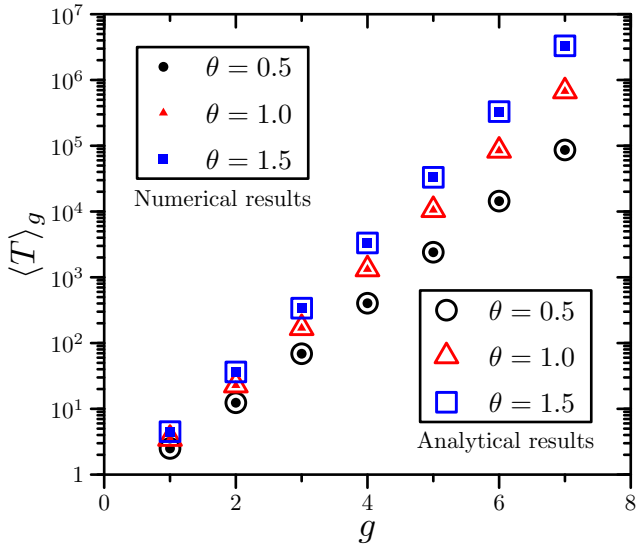}
\end{center}
\caption{MFPT $\langle T \rangle_g$ as a function of $g$ for different networks with various $\theta$. The filled symbols are the data obtained by direct calculation from Eq.~(\ref{Form05}); while the empty symbols are those  exact analytical values given by Eq.~(\ref{FPTF18}).}
\label{compare}
\end{figure}

We have checked the analytical solution in Eq.~(\ref{FPTF18}) against extensive numerical results obtained from Eq.~(\ref{Form05}), see Fig.~\ref{compare}. For different $\theta$ and $g$, both the analytical and numerical results are in full agreement  with each other, indicating that the explicit expression in Eq.~(\ref{FPTF18}) is correct. In addition,  for the particular case  $\theta=1$, the network $\vec{F}_g$ is reduced to $F_g$, and Eq.~(\ref{FPTF18}) recovers the result~\cite{ZhXiZhGaGu09} previously obtained for $F_g$. This also validates Eq.~(\ref{FPTF18}).

%%%%%%%%%%%%%%%%%%%%%%%%%%%%%%%%%%%%%%%%%%%%%%%%%%%%%%%%%
% Figure  3
%%%%%%%%%%%%%%%%%%%%%%%%%%%%%%%%%%%%%%%%%%%%%%%%%%%%%%%%%%
%%%%%%%%%%%%%%%%%%%%%%%%%%%%%%%%%%%%%%%%%%%%%%%%%%%%%%%%%%

We proceed to express $\langle T \rangle_g$ in terms of the network size $N_{g}$, in order to uncover how $\langle T \rangle_g$ scales with $N_{g}$. From $N_{g}=4^{g}+1$, we have
$g=\log_{4}(N_{g}-1)$. Then,
\begin{eqnarray}\label{FPTF19}
 \langle T
\rangle_g
&=& \frac{12\theta^3+17\theta^2+7\theta}{8\theta(\theta+1)(4\theta+3)}(N_{g}-1)^{1+\log_{4}(\theta+1)}  \notag\\
&\quad& +\frac{16\theta^3+28\theta^2+20\theta+6}{8\theta(\theta+1)(4\theta+3)}(N_{g}-1)^{\log_{4}(\theta+1)} \notag\\
&\quad& -\frac{3(\theta+1)(2\theta+1)}{4\theta(\theta+1)(4\theta+3)}\,.
\end{eqnarray}
For a very large network (i.e., $N_g \rightarrow \infty$), the leading term of $\langle T \rangle_g$ can be represented as:
\begin{equation}\label{FPTF20}
\langle T \rangle_{g}\sim (N_{g})^{1+\log_{4}(\theta+1)}\,.
\end{equation}
Equation~(\ref{FPTF20}) shows that for the  directed weighted network $\vec{F}_g$, the
MFPT $\langle T \rangle_g$ behaves as a power-law function of the network size $N_{g}$, with the
exponent $\eta(\theta)=1+\log_{4}(\theta+1)$ increasing with the weight parameter $\theta$. Thus, the weight reciprocity has an essential effect on the efficiency on the trapping problem, measured by the MFPT.

\section{Eigenvalues of the fundamental matrix}

We now study the eigenvalues of the fundamental matrix $K_n$ of the trapping problem addressed above. We will determine all the eigenvalues of the fundamental matrix as  well as their multiplicities. Moreover, we will show that the largest eigenvalue has the same leading scaling as that of $\langle T\rangle_g$. To attain this goal, we introduce matrix $P_g$ defined by $P_g=K_g^{-1}$. Let $\lambda_i(g)$ and $\sigma_i(g)$, where $i=1,2,\ldots,N_n-1$, denote the eigenvalues of $P_g$ and $K_g$, such that $\lambda_1(g)\leq \lambda_2(g)\leq \lambda_3(g) \ldots \leq\lambda_{N_g-1}(g)$ and  $\sigma_1(g)\geq \sigma_2(g) \geq  \sigma_3(g) \geq \ldots \geq\sigma_{N_n-1}(g)$. Then, the one-to-one relation $\lambda_i(g)=1/\sigma_i(g)$ holds. Thus, to compute the eigenvalues of matrix $K_n$, we can alternatively determine the eigenvalues for $P_g$.
In the sequel, we will use the decimation method~\cite{DoAlBeKa83,BlVoJuKo04} to find all the eigenvalues of matrix $P_g$.

\subsection{Full spectrum of fundamental matrix}

The decimation procedure~\cite{DoAlBeKa83,BlVoJuKo04} makes it possible to obtain the eigenvalues for related matrix of current iteration from those of the previous iteration.

We now consider the eigenvalue problem for matrix $P_{g+1}$. Let $\alpha$ denote the set of nodes
in network $\vec{F}_g$, and $\beta$ the set of nodes created at iteration $g+1$. Suppose that
$\lambda_{i}(g+1)$ is an eigenvalue of $P_{g+1}$, and $u=(u_{\alpha},u_{\beta})^\top$ is an
eigenvector associated with $\lambda_{i}(g+1)$, where $u_{\alpha}$ and $u_{\beta}$ correspond to nodes belonging to sets
$\alpha$ and $\beta$, respectively. Then, eigenvalue equation for matrix $P_{g+1}$ can be
represented in a block form:
\begin{equation}\label{T2}
\left[\begin{array}{cccc}
 P_{\alpha,\alpha} & P_{\alpha, \beta} \\
 P_{\beta,\alpha} & P_{\beta, \beta}
\end{array}
\right] \left[\begin{array}{cccc}
 u_{\alpha} \\
 u_{\beta}
\end{array}
\right]=\lambda_{i}(g+1) \left[\begin{array}{cccc}
 u_{\alpha} \\
 u_{\beta}
\end{array}
\right] ,
\end{equation}
where $P_{\alpha,\alpha}$ and $P_{\beta, \beta}$ are the identity matrix.

Equation~(\ref{T2}) can be expressed as two equations:
\begin{equation}\label{T3}
P_{\alpha,\alpha}u_{\alpha}+P_{\alpha, \beta}u_{\beta}=\lambda_{i}(g+1)u_{\alpha}\,,
\end{equation}
\begin{equation}\label{T4}
P_{\beta,\alpha}u_{\alpha}+P_{\beta, \beta}u_{\beta}=\lambda_{i}(g+1)u_{\beta}\,.
\end{equation}
%For the sake of convenience, in what follows, we use $I$ to represent the identity matrix of approximate order.
Equation~(\ref{T4}) implies
\begin{equation}\label{T4b}
u_{\beta}=\frac{1}{\lambda_{i}(g+1)-1}P_{\beta,\alpha}u_{\alpha}\,,
\end{equation}
provided that $\lambda_{i}(g+1) \neq 1$. Inserting Eq.~(\ref{T4b}) into Eq.~(\ref{T3})
yields
\begin{equation}\label{T5}
P_{\alpha, \beta}P_{\beta,\alpha}u_{\alpha}=[\lambda_{i}(g+1)-1]^2\,u_{\alpha}.
\end{equation}
In this way, we reduce the problem of determining the eigenvalue $\lambda_{i}(g+1)$ for matrix $P_{g+1}$ of order $4^{g+1}$ to finding the eigenvalue problem of matrix $P_{\alpha, \beta}P_{\beta,\alpha}$ with a smaller order $4^{g}$.

We can prove (see Methods) that
\begin{equation}\label{T55}
P_{\alpha, \beta}P_{\beta,\alpha}=I_g-\frac{1}{2\theta+2}P_g,
\end{equation}
where
$I_g$ is the identity matrix of order $4^{g}$, identical to that of $P_{g}$.
Equation~(\ref{T55}) relates the product matrix $P_{\alpha, \beta}P_{\beta,\alpha}$ to matrix $P_g$. Therefore, the eigenvalues of
matrix $P_{g+1}$ can be expressed in terms of those of matrix $P_g$.

We next show how to obtain the eigenvalues of $P_{g+1}$ through the eigenvalues of $P_g$. According to
Eqs.~(\ref{T5}) and~(\ref{T55}), we can derive
\begin{equation}\label{A3}
P_g u_{\alpha}=-(2\theta+2)\left[\lambda^2_i(g+1)-2\lambda_i(g+1)\right]u_{\alpha}\,.
\end{equation}
Hence, if $\lambda_{i}(g)$ is an eigenvalue of $P_g$ associated with eigenvector $u_a$,
Eq.~(\ref{A3}) indicates
\begin{equation}\label{T9}
\lambda_{i}(g)=-(2\theta+2)\left[\lambda_i(g+1)^2-2\lambda_i(g+1)\right] \,.
\end{equation}
%which can be rewritten as
%\begin{equation}\label{T10}
%(2\theta+2)\lambda_i(g+1)^2-(4\theta+4)\lambda_i(g+1)+\lambda_i(g)=0\,.
%\end{equation}
Solving the above quadratic equation in the variable $\lambda_{i}(g+1)$ given by Eq.~(\ref{T9}), one
obtains the two roots:
\begin{equation}\label{T11}
\lambda_{i,1}(g+1)=1-\sqrt{1-\frac{\lambda_{i}(g)}{2\theta+2}} \,,
\end{equation}
and
\begin{equation}\label{T12}
\lambda_{i,2}(g+1)=1+\sqrt{1-\frac{\lambda_{i}(g)}{2\theta+2}} \,.
\end{equation}
Equations~(\ref{T11}) and~(\ref{T12}) relate $\lambda_{i}(g+1)$ to $\lambda_{i}(g)$, with each
eigenvalue $\lambda_{i}(g)$ of $P_{g}$ giving rise two different eigenvalues of $P_{g+1}$. As a matter of fact, all eigenvalues of the $P_{g+1}$ can be obtained by these two recursive relations. In Methods, we determine the multiplicity of each eigenvalue and show that all the eigenvalues can be found by Eqs.~(\ref{T11}) and~(\ref{T12}).

Since there is a one-to-one relation between the eigenvalues of $P_g$ and the fundamental matrix $K_g$, we thus have also found all the  eigenvalues of $K_g$.

\subsection{The largest eigenvalue of fundamental matrix and MFPT\label{MFPT}}

In the above, we have determined all eigenvalues for the inverse $P_g$ of the fundamental matrix $K_g$ and thus all eigenvalues of $K_g$. Here we continue to estimate the greatest eigenvalue $\sigma_{\rm max}(g)$ of the fundamental matrix $K_g$, which  actually equals the reciprocal of the smallest eigenvalue for matrix $P_g$, denoted by $\lambda_{\rm min}(g)$. Below we will show that in a large network the leading behavior of the MFPT $\langle T \rangle_g$ for trapping in $\vec{F}_g$ and the reciprocal of $\lambda_{\rm min}(g)$ is identical, that is, $\langle T \rangle_g \sim 1/\lambda_{\rm min}(g)=\sigma_{\rm max}(g)$.

We begin by providing some useful properties of eigenvalues for matrix $P_g$.
Assume that $\Delta_g$ is the set of the $4^g$ eigenvalues of matrix $P_{g}$, namely,
$\Delta_g=\{\lambda_1(g),\lambda_2(g),\lambda_3(g),\cdots, \lambda_{4^g}(g)\}$.  According to the
above analysis, $\Delta_g$ can be categorized into two subsets $\Delta_g^{(1)}$ and $\Delta_g^{(2)}$
satisfying $\Delta_g=\Delta_g^{(1)} \cup \Delta_g^{(2)}$, where $\Delta_g^{(1)}$ consists of all eigenvalues 1,
while $\Delta_g^{(2)}$ contains the rest eigenvalues. Thus,
\begin{equation}\label{MFPT01}
\Delta_g^{(1)}= \underbrace {\{ 1,1,1,\ldots,1,1\} }_{2\times{{4}^{g - 1}}}\,.
\end{equation}
These $2\times {{4}^{g - 1}}$ eigenvalues are labeled sequentially by $\lambda_{4^{g-1}+1}(g)$,
$\lambda_{4^{g-1}+2}(g)$, $\cdots$, $\lambda_{3\times4^{g-1}}(g)$, since they provide a natural
increasing order of all eigenvalues for $P_g$, as will been shown.

The remaining $2\times 4^{g-1}$ eigenvalues in set $\Delta_g^{(2)}$ are all determined by
Eqs.~(\ref{T11}) and~(\ref{T12}). Let $\lambda_1(g-1)$, $\lambda_2(g-1)$, $\lambda_3(g-1)$,
$\cdots$, $\lambda_{4^{g-1}}(g-1)$ be the $4^{g-1}$ eigenvalues of matrix $P_{g-1}$,
arranged in an increasing order $\lambda_1(g-1)\leq \lambda_2(g-1) \leq \lambda_3(g-1) \leq \ldots
\leq \lambda_{4^{g-1}}(g-1)$. Then, for each eigenvalue $\lambda_i(g-1)$ in $P_{g-1}$,
Eqs.~(\ref{T11}) and~(\ref{T12}) produce two eigenvalues of $P_{g}$, which are labeled as $\lambda_{i}(g)$ and
$\lambda_{4^{g}-i+1}(g)$:
\begin{equation}\label{MFPT02}
\lambda_{i}(g)= 1-\sqrt{1-\frac{\lambda_{i}(g-1)}{2\theta+2}}
\end{equation}
and
\begin{equation}\label{MFPT03}
\lambda_{4^g-i+1}(g)=1+\sqrt{1-\frac{\lambda_{i}(g-1)}{2\theta+2}} \,.
\end{equation}
Plugging each eigenvalue of $P_{g - 1}$ into Eqs.~(\ref{T11}) and~(\ref{T12}) generates all
eigenvalues in $\Delta_g^{(2)}$.

It is easy to see that $\lambda_{i}(g)$ given by
Eq.~(\ref{MFPT02}) monotonously increases with $\lambda_i(g-1)$ and belongs to
interval $(0,1)$, while $\lambda_{4^{g}-i+1}(g)$ provided by Eq.~(\ref{MFPT03}) monotonously decreases with $\lambda_i(g-1)$ and lies in interval $(1,2)$. Thus, $\lambda_1(g),\lambda_2(g),\lambda_3(g),\cdots,
\lambda_{4^{g}}(g)$ provide an increasing order of all eigenvalues for matrix $P_{g}$.

We continue to estimate $\lambda_{\rm min}(g)$ of matrix $P_{g}$. From the above arguments, the smallest eigenvalue $\lambda_{\rm min}(g)$ must be the one generated from $\lambda_{\rm
min}(g-1)$ through Eq.~(\ref{MFPT02}):
\begin{equation}\label{MFPT04}
\lambda_{\rm min}(g)= 1-\sqrt{1-\frac{\lambda_{\rm min}(g-1)}{2\theta+2}}\,.
\end{equation}
Using Taylor's formula, we have
\begin{equation}\label{MFPT05}
\lambda_{\rm min}(g) \approx 1-\left[1-\frac{\lambda_{\rm
min}(g-1)}{4\theta+4}\right]=\frac{\lambda_{\rm min}(g-1)}{4\theta+4}\,.
\end{equation}
Considering $\lambda_{\rm min}(1)=1-\sqrt{1-\frac{1}{\theta+1}}$, Eq.~(\ref{MFPT05}) is solved to yield
\begin{equation}\label{MFPT06}
\lambda_{\rm min}(g) \approx \left( 1-\sqrt{1-\frac{1}{\theta+1}} \right)(4\theta+4)^{1-g}\,.
\end{equation}
Thus,
\begin{equation}\label{MFPT07}
\frac{1}{\lambda_{\rm min}(g)}\approx\frac{1}{4} (\theta+1)\left( 1+\sqrt{1+\frac{1}{\theta+1}} \right)4^g(\theta+1)^{g}\,,
\end{equation}
which, together with Eq.~(\ref{FPTF18}), means that $\frac{1}{\lambda_{\rm min}(g)}$ and $\langle T \rangle_g$ have the same dominating term and thus identical leading scaling.

\section{Conclusions}

Real-life weighted networks exhibit a rich and diverse reciprocity structure. In this paper, we have proposed a scale-free weighted directed network with asymmetric edge weights, which are controlled by a parameter characterizing the network reciprocity. We then studied random walks performed on the network with a trap fixed at the central hub node. Applying two different approaches, we have evaluated the MFPT to the trap. Moreover, based on the self-similar architecture of the network, we have found all the eigenvalues and their multiplicities of the fundamental matrix describing the random-walk process, the largest one of which has the same leading scaling as that of the MFPT. The obtained results indicate that  the MFPT scales as a power-law function of the  the system size, with the power exponent increasing with the weight parameter, revealing that the reciprocity has a significant impact on dynamical processes running on weighted networks. This work deepens the understanding of random-walk dynamics in complex systems and opens a novel avenue to control random walks in a weighted network by changing its reciprocity.

\begin{acknowledgments}
The authors thank Bin Wu for his assistance in preparing this manuscript. This work was supported by the National Natural Science Foundation
of China under Grant No. 11275049.
\end{acknowledgments}

\appendix

\section{Proof of Eq.~(\ref{T55}) \label{AppA}}

In order to prove Eq.~(\ref{T55}), we rewrite
$P_{\alpha,\beta}$ and $P_{\beta,\alpha}$ in the block form as
\begin{equation}\label{App01}
P_{\alpha,\beta}=(U_1,U_2,\cdots,U_{E_g})\,
\end{equation}
and
\begin{equation}\label{App02}
P_{\beta,\alpha}=\left(
\begin{array}{c}
V_1 \\
V_2 \\
\vdots \\
V_{E_g} \\
\end{array}\right)\,,
\end{equation}
respectively. In Eqs.~(\ref{App01}) and~(\ref{App02}), $E_g=4^g$ is the number of edges in $F_g$;
$U_i$ ($1 \leq i \leq E_g$) is a $4^g \times 3$ matrix  describing the transition
probability from the $4^g$ non-trap nodes of $F_g$ to the three nodes newly generated by the $i$th edge of $F_g$; similarly,
$V_i$ ($1 \leq i \leq E_g$) is a $3\times 4^g$ matrix indicating the transition
probability from the three new nodes created by the $i$th edge to those $4^g$ old non-trap nodes belonging to
$F_g$. Then,
%\begin{small}
\begin{eqnarray}\label{HA1}
&\quad&P_{\alpha, \beta}P_{\beta,
\alpha}=\sum_{i=1}^{E_g} U_i\, V_i \nonumber\\
&=&\sum_{i=1}^{E_g} \left( \frac{a_i}{\theta+1}\varepsilon_{l_i} + \frac{b_i}{\theta+1}\varepsilon_{r_i},
  \frac{\theta}{\theta+1}a_i\varepsilon_{l_i},
  \frac{\theta}{\theta+1}b_i\varepsilon_{r_i} \right)\times \nonumber \\
  && \quad \quad
  \left(\begin{array}{c}
  -\frac{\varepsilon_{l_i}^{\top}+\varepsilon_{r_i}^{\top}}{2} \\
  \\
  -\varepsilon_{l_i}^{\top} \\
  \\
  -\varepsilon_{r_i}^{\top} \\
  \\
\end{array}\right)\nonumber\\
&=&-\frac{1}{2\theta+2}\times \nonumber\\
&& \sum_{i=1}^{E_g} \left[(2\theta+1)(a_i \varepsilon_{l_i} \varepsilon_{l_i}^{\top}+ b_i \varepsilon_{r_i} \varepsilon_{r_i}^{\top})+
a_i \varepsilon_{l_i} \varepsilon_{r_i}^{\top}
+ b_i \varepsilon_{r_i} \varepsilon_{l_i}^{\top} \right]\nonumber\\
&=&I_g-\frac{1}{2\theta+2}P_g\,,
\end{eqnarray}
%\end{small}
which completes the proof of Eq.~(\ref{T55}).
Note that in Eq.~(\ref{HA1}), $l_i$ and $r_i$ are the two endpoints of the $i$th edge of $F_g$; $\varepsilon_i$ is
a vector having only one nonzero element $1$ at $i$th entry with other entries being zeros; $a_i$ and $b_i$ are two entries of
$P_g$ corresponding to edges $(l_i,r_i)$ and $(r_i,l_i)$, respectively.

\section{Alternative proof of Eq.~(\ref{T55}) \label{AppB}}

Equation~(\ref{T55}) can also be proved using another technique. Assume that $R_g=P_{\alpha,\beta}P_{\beta,\alpha}$ and $Q_g=I_{g}-\frac{1}{2\theta+2}P_{g}$. In order to prove $P_{\alpha,\beta}P_{\beta,\alpha}=I_{g}-\frac{1}{2\theta+2}P_{g}$, it suffices to show that the entries of $R_g$ are equal to their counterparts of $Q_g$. For matrix $Q_g$, it is easy to see that its entries are:  $Q_g(i,i)=\frac{2\theta+1}{2\theta+2}$ for $i= j$ and $Q_g(i,j)=-\frac{1}{2\theta+2}P_g(i,j)$ otherwise. If $P_{g+1}(i,j)$ denotes the $(i,j)$ entry of matrix $P_{g+1}$, the entries of $R_{g}(i,j)$ of matrix $R_g$ can be evaluated by distinguishing two cases: $i=j$ and $i\neq j$.

For the case of $i=j$, the diagonal element of $R_g$ is
\begin{eqnarray}\label{AppB2}
&\quad&R_g(i,i)= \nonumber \\
&=& \displaystyle \sum_{z \in \beta} P_{g+1}(i,z)P_{g+1}(z,i)=\displaystyle \sum_{z \in \beta} \frac{W_{iz}(g+1)}{s^{+}_i(g+1)}\frac{W_{z i}(g+1)}{s^{+}_z(g+1)} \nonumber \\
&=& \frac{1}{2}\sum_{\substack{z \in \beta, i\thicksim z \\ k_z(g+1) = 2 }}\frac{W_{iz}(g+1)}{s^{+}_i(g+1)} + \sum_{\substack{z \in \beta, i\thicksim z \\ k_z(g+1) = 1 }}\frac{W_{iz}(g+1)}{s^{+}_i(g+1)} \nonumber \\
&=& \frac{1}{2}\frac{s^{+}_i(g)}{s^{+}_i(g+1)}+ \frac{\theta \,s^{+}_i(g)}{s^{+}_i(g+1)}\nonumber \\
&=& \frac{2\theta+1}{2\theta + 2}=Q_g(i,i),
\end{eqnarray}
where the relation $s^{+}_i(g+1)=(\theta+1)\,s^{+}_i(g)$ is used. In Eq.~(\ref{AppB2}), $i\thicksim z$  indicates that two nodes $i$ and $z$ are adjacent  in network $F_{g+1}$.

For the other case of $i \neq j$, the non-diagonal element of $R_g$ is
\begin{eqnarray}\label{AppB3}
R_g(i,j)&=&  \sum_{z \in \beta} P_{g+1}(i,z)P_{g+1}(z,j) \nonumber \\
&=& \sum_{\substack{A_{g+1}(i,z) = 1 \\ A_{g+1}(z,j) = 1 }} \frac{W_{iz}(g+1)}{s^{+}_i(g+1)}\frac{W_{z j}(g+1)}{s^{+}_z(g+1)} \nonumber \\
&=&\frac{1}{2} \frac{W_{ij}(g)}{s^{+}_i(g+1)}=-\frac{1}{2\theta+2}P_g(i,j)\nonumber \\
&=&Q_g(i,j),
\end{eqnarray}
which, together with~(\ref{AppB2}) proves  Eq.~(\ref{T55}).

\section{Multiplicities of eigenvalues \label{AppB}}

By numerically computing the eigenvalues for the first several iterations, we can observe some important phenomena and properties about the structure of the
eigenvalues. When $g=1$, the eigenvalues of $P_1$ are $1-\sqrt{1-\frac{1}{\theta+1}}$ and $1+\sqrt{1-\frac{1}{\theta+1}}$, both of which have a multiplicity of $2$. When $g=2$, $P_2$ have 16 eigenvalues:  eigenvalue 1 with degeneracy 8 and 4 two-fold other eigenvalues generated by $1-\sqrt{1-\frac{1}{\theta+1}}$ and $1+\sqrt{1-\frac{1}{\theta+1}}$. When $g \geq 3$, all the eigenvalues $P_g$ can be put into two classes. The first class includes  eigenvalue 1 and those generated by $1$, which display the following feature that each eigenvalue appearing at a given iteration $g_{i}$ will continue to appear at all subsequent generations greater than $g_{i}$.
The second class contains those eigenvalues generated by the two $1-\sqrt{1-\frac{1}{\theta+1}}$ and $1+\sqrt{1-\frac{1}{\theta+1}}$ in $P_1$. Each eigenvalue in this class is two-fold, and  each eigenvalue of a given iteration $g_{i}$ does not appear at any of subsequent iterations larger than $g_{i}$. For the two eigenvalue classes,  each eigenvalue (other than $1$) of current generation keeps the multiplicity of its father of the previous generation.

%All new eigenvalues appearing at iteration $g_{i}+1$ are just those generated via Eqs.~(\ref{T11})  and~(\ref{T12}) by substituting $\lambda_{i}(g)$ with $\lambda_{i}(g_{i})$ that are newly created at iteration $g_{i}$.

Using the above-observed properties of the eigenvalue structure, we can determine the multiplicities of all eigenvalues. Let $M_g(\lambda)$ denote the multiplicity of eigenvalue $\lambda$ of matrix $P_g$. We first determine the number of eigenvalue $1$ for
$P_g$. To this end, let $r(X)$ denote the rank of matrix $X$. Then
\begin{equation}\label{N0}
M_g(\lambda=1)=4^g-r(P_g-1\times I_g)\,.
\end{equation}
For $g=1$, $M_1(1)=0$; for $g=1$, $M_2(1)=8$. For $g\geq 2$, it is obvious that $r(P_{g+1}-I_{g+1})=r(P_{\alpha,\beta}) + r(P_{\beta,\alpha})$, where $r(P_{\alpha,\beta})$ and $r(P_{\beta,\alpha})$ can be determined in the following way.

We first show that $P_{\beta,\alpha}$ is a full column rank matrix. Let
\begin{equation}
\phi=(\phi_1,\phi_2,\cdots ,\phi_{3\times 4^g})^{\top} = \sum_{\substack{i\in \alpha \\ i \neq 1}} k_iM_i,
\end{equation}
where $M_i$ is the column vector of $P_{\beta,\alpha}$ representing the $i$th column of $P_{\beta,\alpha}$. Let $M_i = (M_{1,i},M_{2,i},\cdots,M_{3\times 4^g,i})^{\top}$. Suppose that $\phi=0$. Then, we can prove that for an arbitrary $k_i$, $k_i=0$ always holds. By construction, for any old node $i \in \alpha$, there exists a new leaf node $l \in \beta$ attached to $i$. Then, for $\phi_l=k_1M_{1,l}+k_2M_{2,l},\cdots,+k_{3\times 4^g}M_{3\times 4^g,l}$, only $M_{i,l}\neq 0$ but all $M_{x,l}=0$ for $x\neq i$. From $\phi_l=0$, we have $k_i=0$. Therefore, $r(P_{\beta,\alpha})=4^g$. Analogously, we can verify that $P_{\alpha,\beta}$ is  a full row rank matrix and $r(P_{\alpha,\beta})=4^g$.

%We next consider a comparison between $P_{\alpha,\beta}$ and $P_{\beta,\alpha}^{\top}$. For every entry, it's either zero in both matrices or nonzero in both matrices, so in the same way we can prove that $P_{\alpha,\beta}$ is a full row rank matrix. Therefore, $r(P_{\alpha,\beta})=E_g=4^g$.

Combining the above results,  the multiplicity  of eigenvalue 1 of $P_g$ is
\begin{equation}\label{N4}
M_g(\lambda=1)=\begin{cases}
0, &g=1, \\
2\times 4^{g-1}, &g \geqslant 2.
\end{cases}
\end{equation}
We continue to compute the multiplicities of other eigenvalues generated by $1$ that are in the first eigenvalue class. Since every eigenvalue at a given iteration keeps the multiplicity of its father at the preceding iteration, for matrix $P_g$, the multiplicity of each first-generation descendant of
eigenvalue 1 is $2\times 4^{g-2}$, the multiplicity of each second-generation descendant of eigenvalue
1 is $2\times 4^{g-3}$, and the multiplicity of each $(g-2)$nd generation descendant of eigenvalue 1 is
$2\times4$. Moreover, we can derive that that the number of the $i$th ($0\leq i\leq g-2$) generation distinct descendants of eigenvalue is $2^{i}$, where $0$th generation descendants refer to the $2\times 4^{g-1}$ eigenvalues $1$ themselves. Finally, it is easy to verify that the number of all the eigenvalues in the second eigenvalue class is $4\times 2^{g-1}$.
Hence, the total number of  eigenvalues of matrix $P_g$ is
\begin{equation}
\sum_{i=0}^{g-2}\left[\left(2\times 4^{g-1-i}\right)\times 2^{i}\right]+4\times 2^{g-1}= 4^g\,,
\end{equation}
indicating that all the eigenvalues of $P_g$ are successfully found.

%\nocite{*}
%\bibliography{aipsamp}% Produces the bibliography via BibTeX.

\end{document}